\documentclass[aos,preprint]{imsart}
\RequirePackage[OT1]{fontenc}
\RequirePackage{amsthm,amsmath}
\RequirePackage[numbers]{natbib}
\usepackage{amsmath}
\usepackage{latexsym, epsfig, amssymb, amsmath, amsthm, graphicx, mathrsfs, booktabs}
\usepackage{bm}
\usepackage{lscape}
\usepackage{subcaption}
\usepackage{graphbox}
\usepackage{float}

\startlocaldefs

\theoremstyle{plain}
\newtheorem{theorem}{Theorem}
\newtheorem{assumption}{Condition}

\newcommand{\aphi}{\phi}
\newcommand{\ba}{\mbox{\bf a}}
\newcommand{\bb}{\mbox{\bf b}}
\newcommand{\be}{\mbox{\bf e}}

\newcommand{\br}{\mbox{\bf r}}

\newcommand{\bu}{\mbox{\bf u}}

\newcommand{\bx}{\mbox{\bf x}}
\newcommand{\by}{\mbox{\bf y}}
\newcommand{\bz}{\mbox{\bf z}}
\newcommand{\bA}{\mbox{\bf A}}

\newcommand{\bQ}{\mbox{\bf Q}}

\newcommand{\bX}{\mbox{\bf X}}

\newcommand{\bY}{\mbox{\bf Y}}

\newcommand{\bone}{\mbox{\bf 1}}
\newcommand{\bzero}{\mbox{\bf 0}}

\newcommand{\bbeta}{\mbox{\boldmath $\beta$}}

\newcommand{\bdelta}{\mbox{\boldmath $\delta$}}
\newcommand{\btheta}{\mbox{\boldmath $\theta$}}

\newcommand{\bgamma}{\mbox{\boldmath $\gamma$}}
\newcommand{\bPsi}{\mbox{\boldmath $\Psi$}}
\newcommand{\bet}{\mbox{\boldmath $\eta$}}
\newcommand{\bxi}{\mbox{\boldmath $\xi$}}

\newcommand{\bmu}{\mbox{\boldmath $\mu$}}

\newcommand{\hbbeta}{\widehat\bbeta}

\newcommand{\hbeta}{\widehat{\beta}}

\newcommand{\var}{\mathrm{var}}

\newcommand{\Sig}{\mathbf{\Sigma}}

\newcommand{\diag}{\mathrm{diag}}

\def\t{^T}
\def\toD{\overset{\mathscr{D}}{\longrightarrow}}

\def\eqd{\, {\buildrel d \over =}\, }

\endlocaldefs

\begin{document}

\begin{frontmatter}
\title{Nonuniformity of P-values Can Occur Early in Diverging Dimensions \thanksref{T1}}
\runtitle{Nonuniformity of P-values}
\thankstext{T1}{
This work was supported by NSF CAREER Award DMS-1150318 and a grant from the Simons Foundation. The first and last authors sincerely thank Emmanuel Cand\`es for helpful discussions on this topic.}

\begin{aug}
\author{\fnms{Yingying} \snm{Fan}\ead[label=e1]{fanyingy@marshall.usc.edu}},
\author{\fnms{Emre} \snm{Demirkaya}\ead[label=e2]{demirkay@usc.edu}}
\and
\author{\fnms{Jinchi} \snm{Lv}\ead[label=e3]{jinchilv@marshall.usc.edu}}

\runauthor{Y. Fan, E. Demirkaya and J. Lv}

\affiliation{University of Southern California}

\address{Data Sciences and Operations Department\\
Marshall School of Business\\
University of Southern California\\
Los Angeles, CA 90089\\
USA\\
\printead{e1}\\
\phantom{E-mail:\ }\printead*{e3}}

\address{Department of Mathematics\\
University of Southern California\\
Los Angeles, CA 90089\\
USA\\
\printead{e2}}
\end{aug}

\begin{abstract}
Evaluating the joint significance of covariates is of fundamental importance in a wide range of applications. To this end, p-values are frequently employed and produced by algorithms that are powered by classical large-sample asymptotic theory. It is well known that the conventional p-values in Gaussian linear model are valid even when the dimensionality is a non-vanishing fraction of the sample size, but can break down when the design matrix becomes singular in higher dimensions or when the error distribution deviates from Gaussianity. A natural question is when the conventional p-values in generalized linear models become invalid in diverging dimensions. We establish that such a breakdown can occur early in nonlinear models. Our theoretical characterizations are confirmed by simulation studies.
\end{abstract}



\end{frontmatter}

\section{Introduction} \label{Sec1}
In many applications it is often desirable to evaluate the significance of covariates in a predictive model for some response of interest. Identifying a set of significant covariates can facilitate domain experts to further probe their causal relationships with the response. Ruling out insignificant covariates can also help reduce the fraction of false discoveries and narrow down the scope of follow-up experimental studies by scientists. These tasks certainly require an accurate measure of feature significance in finite samples. The tool of p-values has provided a powerful framework for such investigations.

As p-values are routinely produced by algorithms, practitioners should perhaps be aware that those p-values are usually based on classical large-sample asymptotic theory. For example, marginal p-values have been employed frequently in large-scale applications when the number of covariates $p$ greatly exceeds the number of observations $n$. Those p-values are based on marginal regression models linking each individual covariate to the response separately. In these marginal regression models, the ratio of sample size to model dimensionality is equal to $n$, which results in justified p-values as sample size increases. Yet due to the correlations among the covariates, we often would like to investigate the joint significance of a covariate in a regression model conditional on all other covariates, which is the main focus of this paper. A natural question is whether conventional joint p-values continue to be valid in the regime of diverging dimensionality $p$.

It is well known that fitting the linear regression model with $p > n$ using the ordinary least squares can lead to perfect fit giving rise to zero residual vector, which renders the p-values undefined. When $p \leq n$ and the design matrix is nonsingular, the p-values in the linear regression model are well defined and valid thanks to the exact normality of the least-squares estimator when the random error is Gaussian and the design matrix is deterministic. When the error is non-Gaussian, \cite{Huber1973} showed that the least-squares estimator can still be asymptotically normal under the assumption of $p = o(n)$, but is generally no longer normal when $p = o(n)$ fails to hold, making the conventional p-values inaccurate in higher dimensions. For the asymptotic properties of $M$-estimators for robust regression, see, for example, \cite{Huber1973, Portnoy1984, Portnoy1985} for the case of diverging dimensionality $p = o(n)$ and \cite{KarouiBeanBickel13, BeanDerekBickel13} for the scenario when the dimensionality $p$ grows proportionally to sample size $n$.

We have seen that the conventional p-values for the least-squares estimator in linear regression model can start behaving wildly and become invalid when the dimensionality $p$ is of the same order as sample size $n$ and the error distribution deviates from Gaussianity. A natural question is whether similar phenomenon holds for the conventional p-values for the maximum likelihood estimator (MLE) in the setting of diverging-dimensional nonlinear models. More specifically, we aim to answer the question of whether $p\sim n$ is still the breakdown point of the conventional p-values when we move away from the regime of linear regression model, where $\sim$ stands for asymptotic order.  To simplify the technical presentation, in this paper we adopt the generalized linear model (GLM) as a specific family of nonlinear models \cite{MN89}. The GLM with a canonical link assumes that the conditional distribution of $\by$
given $\bX$ belongs to the canonical exponential family, having
the following density function with respect to some fixed measure
\begin{eqnarray} \label{001}
f_n(\by; \bX, \bbeta) \equiv  \prod_{i = 1}^n f_0(y_i; \theta_i)  =  \prod_{i = 1}^n \left\{c(y_i) \exp\left[\frac{y_i \theta_i - b(\theta_i)}{\aphi}\right]\right\},
\end{eqnarray}
where $\bX = (\bx_1, \cdots, \bx_p)$ is an $n \times p$ design matrix with $\bx_j = (x_{1j}, \cdots, x_{nj})\t$, $j = 1, \cdots, p$, $\by = (y_1, \cdots, y_n)\t$ is an $n$-dimensional response vector, $\bbeta = (\beta_1, \cdots, \beta_p)\t$ is a
$p$-dimensional regression coefficient vector, $\{f_0(y; \theta): \theta \in \mathbb{R}\}$ is a family of distributions in
the regular exponential family with dispersion parameter $\aphi \in (0, \infty)$, and
$\btheta = (\theta_1, \cdots, \theta_n)\t = \bX \bbeta$. As is common in GLM,
the function $b(\theta)$ in (\ref{001}) is implicitly assumed to be twice
continuously differentiable with $b''(\theta)$ always positive. Popularly used GLMs include the linear regression model, logistic regression model, and Poisson regression model for continuous, binary, and count data of responses, respectively.

The key innovation of our paper is the formal justification that the conventional p-values in nonlinear models of GLMs can become invalid in diverging dimensions and such a breakdown can occur \textit{much earlier} than in linear models, which spells out a fundamental difference between linear models and nonlinear models. To begin the journey of p-values in diverging-dimensional GLMs, let us gain some insights into this problem by looking at the specific case of logistic regression. Recently, \cite{Candes2016} established an interesting phase transition phenomenon of perfect hyperplane separation for high-dimensional classification with an elegant probabilistic argument. Suppose we are given a random design matrix $\bX \sim N(\bzero, I_n\otimes I_p)$ and arbitrary binary $y_i$'s that are not all the same. The phase transition of perfect hyperplane separation happens at the point $p/n = 1/2$. With such a separating hyperplane, there exist some  $\bbeta^* \in \mathbb{R}^p$ and $t \in \mathbb{R}$ such that $\bx_i\t \bbeta^* > t$ for all cases $y_i = 1$ and $\bx_i\t \bbeta^* < t$ for all controls $y_i = 0$. Let us fit a logistic regression model with an intercept. 
It is easy to show that multiplying the vector $(-t, (\bbeta^*)\t)\t$ by a divergence sequence of positive numbers $c$, we can obtain a sequence of logistic regression fits with the fitted response vector approaching $\by = (y_1, \cdots, y_n)\t$
as $c \rightarrow \infty$. As a consequence, the MLE
algorithm can return a pretty wild estimate that is close to infinity in topology when the algorithm is set to stop.
Clearly, in such a case the p-value of the MLE is no longer justified and meaningful. The results in \cite{Candes2016} motivate us to characterize the 
breakdown point of p-values in nonlinear GLMs with $p \sim n^{\alpha_0}$ in the regime of $\alpha_0 \in [0,1)$.

It is worth mentioning that our work is different in goals from the limited but growing literature on p-values for high-dimensional nonlinear models, and makes novel contributions to such a problem. The key distinction is that existing work has focused primarily on identifying the scenarios in which conventional p-values or their modifications continue to be valid with some sparsity assumption limiting the growth of intrinsic dimensions. For example, \cite{FP04} established the oracle property including the asymptotic normality for nonconcave penalized likelihood estimators in the scenario of $p = o(n^{1/5})$, while \cite{FanLv2011} extended their results to the GLM setting of non-polynomial (NP)
dimensionality. In the latter work, the p-values were proved to be valid under the assumption that the intrinsic dimensionality $s = o(n^{1/3})$. More recent work on high-dimensional inference in nonlinear model settings includes \cite{vandeGeeretal2014, Atheyetal2016} under sparsity assumptions. In addition, two tests were introduced in \cite{GuoChen2016} for high-dimensional GLMs without or with nuisance regression parameters, but the p-values were obtained for testing the global hypothesis for a given set of covariates, which is different from our goal of testing the significance of individual covariates simultaneously.

The rest of the paper is organized as follows. Section \ref{Sec2} provides characterizations of p-values in low dimensions. We establish the nonuniformity of GLM p-values in diverging dimensions in Section \ref{Sec3}. Section \ref{Sec4} presents several simulation examples verifying the theoretical phenomenon. We discuss some implications of our results in Section \ref{Sec5}. The proofs of all the results are relegated to the Appendix.

\section{Characterizations of p-values in low dimensions} \label{Sec2}
To pinpoint the 
breakdown point of GLM p-values in diverging dimensions, we start with characterizing p-values in low dimensions. In contrast to existing work on the asymptotic distribution of the penalized MLE, our results in this section focus on the asymptotic normality of the unpenalized MLE in diverging-dimensional GLMs, which justifies the validity of conventional p-values. Although Theorem \ref{Thm3} to be established is in the conventional sense, to the best of our knowledge such results are not available in the literature before in terms of the maximum range of the dimensionality $p$ without any sparsity assumption.

\subsection{Maximum likelihood estimation and technical conditions} \label{Sec2.1}
For the GLM (\ref{001}), the log-likelihood $\log f_n(\by; \bX,
\bbeta)$ of the sample is given, up to an affine transformation, by
\begin{equation} \label{002}
\ell_n(\bbeta) = n^{-1}  \left[\by\t \bX \bbeta - \bone\t \bb(\bX
\bbeta)\right],
\end{equation}
where $\bb(\btheta) = (b(\theta_1), \cdots, b(\theta_n))\t$ for $\btheta = (\theta_1, \cdots, \theta_n)\t \in \mathbb{R}^n$. Denote by $\hbbeta = (\hbeta_1, \cdots, \hbeta_p)\t \in \mathbb{R}^p$ the MLE which is the maximizer of (\ref{002}), and
\begin{equation} \label{105}
\bmu(\btheta)  = (b'(\theta_1), \cdots, b'(\theta_n))\t \ \text{ and } \
\Sig(\btheta)  =  \diag\{b''(\theta_1), \cdots, b''(\theta_n)\}.
\end{equation}
A well-known fact is that the $n$-dimensional response vector $\by$ in GLM (\ref{001}) has mean vector $\bmu(\btheta)$ and
covariance matrix $\aphi \Sig(\btheta)$. Clearly, the MLE $\hbbeta$ is given by the unique solution to the score equation
\begin{equation} \label{021}
\bX\t [\by - \bmu(\bX \bbeta)] = \bzero
\end{equation}
when the design matrix $\bX$ is of full column rank $p$.

We first introduce a deviation probability bound that facilitates our technical analysis.
Consider both cases of bounded responses and unbounded responses. 
In the latter case, assume that there exist some constants $M, v_0 > 0$ such that
\begin{equation} \label{010}
\max_{1 \leq i \leq n} E \left\{\exp\left[\frac{\left|y_i - b'\left(\theta_{0, i}\right)\right|}{M}\right] - 1 - \frac{\left|y_i - b'\left(\theta_{0, i}\right)\right|}{M}\right\} M^2 \leq \frac{v_0}{2}
\end{equation}
with $(\theta_{0, 1}, \cdots, \theta_{0, n})\t = \btheta_0 = \bX \bbeta_0$, where $\bbeta_0 = (\beta_{0, 1}, \cdots,
\beta_{0, p})\t$ denotes the true
regression coefficient vector in model (\ref{001}). Then by \cite{FanLv2011, FanLv2013}, it holds that for any $\ba \in \mathbb{R}^n$,
\begin{equation} \label{137}
P\left(\left|\ba\t \bY - \ba\t \bmu\left(\btheta_0\right)\right| >
  \left\|\ba\right\|_2 \varepsilon \right ) \leq \varphi(\varepsilon),
\end{equation}
where $\varphi(\varepsilon) = 2 e^{-c_1 \varepsilon^2}$ with $c_1 > 0$ some constant, and $\varepsilon \in (0, \infty)$ if the responses are bounded and $\varepsilon \in (0, \|\ba\|_2 /\|\ba\|_\infty]$ if the responses are unbounded.

For nonlinear GLMs, the MLE $\hbbeta$ solves the nonlinear score equation (\ref{021}) whose solution generally does not admit an explicit form. To address such a challenge, we construct a solution to equation (\ref{021}) in an asymptotically shrinking neighborhood of $\bbeta_0$ that meets the MLE $\hbbeta$ thanks to the uniqueness of the solution. Specifically, define a neighborhood of $\bbeta_0$ as
\begin{equation} \label{004}
\mathcal{N}_0 = \{\bbeta \in \mathbb{R}^p: \|\bbeta -
\bbeta_0\|_\infty \leq n^{-\gamma} \log n\}
\end{equation}
for some constant $\gamma \in (0, 1/2]$. Assume that $p = O(n^{\alpha_0})$ for some $\alpha_0 \in (0, \gamma)$ and let $b_n = o\{\min(n^{1/2 -\gamma} \sqrt{\log n}, n^{2\gamma - \alpha_0 - 1/2}/ (\log n)^2\}$ be a diverging sequence of positive numbers. We need some basic regularity conditions to establish the asymptotic normality of the MLE $\hbbeta$.


\begin{assumption} \label{con1}
The design matrix $\bX$ satisfies
	\begin{align} \label{007}
	& \left\|\left[\bX\t \Sig\left(\btheta_0\right) \bX\right]^{-1}\right\|_\infty
	= O(b_n n^{-1}), \\
	\label{009}
	& \max_{\bbeta \in \mathcal{N}_0} \max\nolimits_{j = 1}^p \lambda_{\max}\left[\bX\t \diag\left\{\left|\bx_j\right| \circ \left|\bmu''\left(\bX \bbeta\right)\right|\right\} \bX\right] = O(n)
	\end{align}
with $\circ$ denoting the Hadamard 
product and derivatives understood componentwise. Assume that $\max_{j = 1}^p \|\bx_j\|_\infty < c_1^{1/2} \{n/(\log n)\}^{1/2}$ if the responses are unbounded.
\end{assumption}

\begin{assumption} \label{con2}
The 
eigenvalues of $n^{-1} \bA_n$ are bounded away from $0$ and $\infty$, $\sum_{i = 1}^n (\bz_i\t \bA_n^{-1} \bz_i)^{3/2} = o(1)$, and $\max_{i = 1}^n E|y_i - b'(\theta_{0, i})|^3 = O(1)$, where $\bA_n = \bX\t \Sig(\btheta_0) \bX$ and $(\bz_1, \cdots, \bz_n)\t = \bX$.
\end{assumption}

Conditions \ref{con1} and \ref{con2} put some basic restrictions on the design matrix $\bX$ and a moment condition on the responses. For the case of linear model, bound \eqref{007} becomes  $\|(\bX\t\bX)^{-1}\|_\infty = O(b_n/n)$ and bound \eqref{009} holds automatically since $b'''(\theta) \equiv 0$. 
Condition \ref{con2}  is related to the Lyapunov condition.

\subsection{Conventional p-values in low dimensions} \label{Sec2.2}

\begin{theorem}[Asymptotic normality] \label{Thm3}
Assume that Conditions \ref{con1}--\ref{con2} and probability bound (\ref{137}) hold. Then the MLE $\hbbeta$ satisfies that for each $1 \leq j \leq p$,
\begin{equation} \label{011}
	(\bA_n^{-1})_{jj}^{-1/2} (\hbeta_j - \beta_{0,j}) \toD N(0, \aphi),
\end{equation}
where $\bA_n = \bX\t \Sig(\btheta_0) \bX$ and $(\bA_n^{-1})_{jj}$ denotes the $j$th diagonal entry.
\end{theorem}

Theorem \ref{Thm3} establishes the asymptotic normality of the MLE and consequently justifies the validity of the conventional p-values in low dimensions. Note that for simplicity, we present here only the marginal asymptotic normality, and the joint asymptotic normality also holds for the projection of the MLE onto any fixed-dimensional subspace. This result can also be extended to the case of misspecified models; see, for example, \cite{LvLiu2014}.


As mentioned in the Introduction, the asymptotic normality was shown in \cite{FanLv2011} for nonconcave penalized MLE having intrinsic dimensionality $s = o(n^{1/3})$. In contrast, our result in Theorem \ref{Thm3} allows for the scenario of $p = o(n^{1/2})$ with no sparsity assumption in view of our technical conditions. In particular, we see that the conventional p-values in nonlinear GLMs generally remain valid in the regime of slowly diverging dimensionality $p = o(n^{1/2})$.


\section{Nonuniformity of GLM p-values in diverging dimensions}  \label{Sec3}
So far we have seen that for nonlinear GLMs, the p-values can be valid when $p = o(n^{1/2})$ as shown in Section \ref{Sec2}, and can become meaningless when $p \geq n/2$ as discussed in the Introduction. Apparently, there is a big gap between these two regimes of growth of dimensionality $p$. To provide some guidance on the practical use of p-values in nonlinear GLMs, it is of crucial importance to characterize their 
breakdown point.
To highlight the main message with simplified technical presentation, hereafter we content ourselves on the specific case of logistic regression model for binary response. We argue that this specific model is sufficient for our purpose because for conventional p-values derived from MLEs in diverging-dimensional GLMs to be valid, it must be at least valid for the specific model of logistic regression. Therefore, the breakdown point for logistic regression is at least the breakdown point for general nonlinear GLMs. This argument is fundamentally different from that of proving the overall validity of conventional p-values, where one needs to prove the asymptotic normality of MLEs under general GLMs rather than any specific model.


\subsection{The wild side of nonlinear regime} \label{Sec3.1}
For the logistic regression model (\ref{001}), we have $b(\theta) = \log(1 +
e^\theta)$, $\theta \in \mathbb{R}$ and $\aphi = 1$. The mean vector $\bmu(\btheta)$ and covariance matrix $\aphi \Sig(\btheta)$ of the $n$-dimensional response vector $\by$ given by (\ref{105}) now take the familiar form of
$\bmu(\btheta) = \left(\frac{e^{\theta_1}}{1 + e^{\theta_1}},
\cdots, \frac{e^{\theta_n}}{1 + e^{\theta_n}}\right)\t$ and
\[ \Sig(\btheta) = \diag\left\{\frac{e^{\theta_1}}{\left(1 + e^{\theta_1}\right)^2}, \cdots, \frac{e^{\theta_n}}{\left(1 + e^{\theta_n}\right)^2}\right\} \]
with $\btheta = (\theta_1, \cdots, \theta_n)\t = \bX \bbeta$. In many real applications, one would like to interpret the significance of each individual covariate produced by algorithms based on the conventional asymptotic normality of the MLE as established in Theorem \ref{Thm3}. As argued at the beginning of this section, in order to justify any significant discoveries, the underlying theory of p-values in diverging-dimensional GLMs should ideally at least ensure that the distributional property in (\ref{011}) holds for the scenario of true regression coefficient vector $\bbeta_0 = \bzero$, that is, under the global null. Otherwise practitioners may simply lose the theoretical backup and the resulting decisions based on the p-values can become ineffective or even misleading. For this reason, we identify the 
breakdown point of p-values in diverging-dimensional logistic regression model under the global null.

Characterizing the 
breakdown point of p-values in nonlinear GLMs is highly nontrivial and challenging. First, the nonlinearity generally renders the MLE to take no analytical form, which makes it difficult to analyze its behavior in diverging dimensions. Second, conventional probabilistic arguments for establishing the central limit theorem of MLE only enable us to see one side of the coin, but not exactly at what point the distributional property fails to hold. To address these important challenges, we introduce novel geometric and probabilistic arguments presented later in the proofs of Theorems \ref{Thm1}--\ref{Thm2} that provide a rather delicate analysis of the MLE. In particular, our arguments unveil that the early breakdown point of p-values in nonlinear GLMs is essentially due to the \textit{nonlinearity} of the mean function $\bmu(\cdot)$. This shows that p-values can behave wildly much early on in diverging dimensions when we travel from the linear regime to the nonlinear world as simple as the widely applied logistic regression; see the Introduction for detailed discussions on the p-values in diverging-dimensional linear models.

Before presenting the main results, let us look at the specific case of logistic regression model under the global null. In such a scenario, it holds that $\btheta_0 = \bX \bbeta_0 = \bzero$ and thus $\Sig(\btheta_0) = 4^{-1} I_n$, which results in
\[ \bA_n = \bX\t \Sig(\btheta_0) \bX = 4^{-1} \bX\t \bX. \]
In particular, we see that when $n^{-1} \bX\t \bX$ is close to the identity matrix $I_p$, the asymptotic standard deviation of the $j$th component $\hbeta_j$ of the MLE $\hbbeta$ is close to $2 n^{-1/2}$ when the asymptotic theory in (\ref{011}) holds. As mentioned in the Introduction, when $p \geq n/2$ the MLE can blow up with excessively large variance, a strong evidence against the distributional property in (\ref{011}). In fact, one can also observe inflated variance of the MLE relative to what is predicted by the asymptotic theory in (\ref{011}) even when the dimensionality $p$ grows at a slower rate with sample size $n$. As a consequence, the conventional p-values given by algorithms according to property (\ref{011}) can be much biased toward zero and thus produce more significant discoveries than the truth. Such a breakdown of conventional p-values is delineated clearly in the simulation examples presented in Section \ref{Sec4}.

\subsection{Main results} \label{Sec3.2}

We now present the formal results on the invalidity of GLM p-values in diverging dimensions.

\begin{theorem}[Uniform orthonormal design] \label{Thm1}
Assume that 
$n^{-1/2} \bX$ is uniformly distributed on the Stiefel manifold $V_p(\mathbb{R}^n)$ consisting of all $n \times p$ orthonormal matrices. Then for the logistic regression model under the global null, the asymptotic normality of the MLE established in (\ref{011}) fails to hold when $p \sim n^{2/3}$, where $\sim$ stands for asymptotic order.
\end{theorem}

\begin{theorem}[Correlated Gaussian design] \label{Thm2}
Assume that $\bX \sim N (\bzero, I_{n} \otimes \Sig )$ with covariance matrix $\Sig$ nonsingular. Then for the logistic regression model under the global null, the same conclusion as in Theorem \ref{Thm1} holds.
\end{theorem}

The key ingredients of our new geometric and probabilistic arguments are demonstrated in the proof of Theorem \ref{Thm1} in Section \ref{SecA.2}. The assumption that the rescaled random design matrix $n^{-1/2} \bX$ has the  Haar measure on the Stiefel manifold $V_p(\mathbb{R}^n)$ greatly facilitates our technical analysis. The major theoretical finding is that the nonlinearity of the mean function $\bmu(\cdot)$ can be negligible in determining the asymptotic distribution of MLE as given in (\ref{011}) when the dimensionality $p$ grows at a slower rate than $n^{2/3}$, but such nonlinearity can become dominant and deform the conventional asymptotic normality when $p$ grows at rate $n^{2/3}$ or faster. See the last paragraph of Section \ref{SecA.2} for more detailed in-depth discussions on such an interesting phenomenon.

Theorem \ref{Thm2} further establishes that the invalidity of GLM p-values in high dimensions beyond the scenario of orthonormal design matrices considered in Theorem \ref{Thm1}. 
The breakdown of the conventional p-values can be regardless of the correlation structure of the covariates.

Our theoretical 
derivations detailed in the Appendix also 
suggest that the conventional p-values in nonlinear GLMs can generally fail to be valid when $p \sim n^{\alpha_0}$ with $\alpha_0$ ranging between $1/2$ and $2/3$, which differs significantly from the phenomenon for linear models as discussed in the Introduction. 
The special feature of logistic regression model that the variance function $b''(\theta)$ takes the maximum value $1/4$ at natural parameter $\theta = 0$ leads to a higher transition point of $p \sim n^{\alpha_0}$ with $\alpha_0 = 2/3$ for the case of global null $\bbeta_0 = \bzero$.

\section{Numerical studies} \label{Sec4}
We now delineate the journey of p-values of nonlinear GLMs in diverging dimensions as predicted by our major theoretical results in Section \ref{Sec3} with several simulation examples. Indeed, these theoretical results are well supported by the numerical studies.


\subsection{Simulation examples} \label{Sec4.1}
Following Theorems \ref{Thm1}--\ref{Thm2} in Section \ref{Sec3}, we consider three examples of the logistic regression model (\ref{001}). The response vector $\by = (y_1, \cdots, y_n)\t$ has independent components and each $y_i$ has Bernoulli distribution with parameter
$e^{\theta_i}/(1 + e^{\theta_i})$, where $\btheta = (\theta_1, \cdots, \theta_n)\t = \bX \bbeta_0$. In example 1, we generate the $n \times p$ design matrix $\bX = (\bx_1, \cdots, \bx_p)$ such that $n^{-1/2} \bX$ is uniformly distributed on the Stiefel manifold $V_p(\mathbb{R}^n)$ as in Theorem \ref{Thm1}, while examples 2 and 3 assume that $\bX \sim N (\bzero, I_{n} \otimes \Sig)$ with covariance matrix $\Sig$ as in Theorem \ref{Thm2}. In particular, we choose $\Sig = (\rho^{|j-k|})_{1 \leq j,k \leq p}$ with $\rho = 0, 0.5$, and $0.8$ to reflect low, moderate, and high correlation levels among the covariates. Moreover, examples 1 and 2 assume the global null model with $\bbeta_0 = \bzero$ following our theoretical results, whereas example 3 allows sparsity $ s = \| \bbeta_{0} \| _{0}$ to vary.

To examine the asymptotic results we set the sample size $n = 1000$. In each example, we consider a spectrum of dimensionality $p$ with varying rate of growth with sample size $n$. As mentioned in the Introduction, the phase transition of perfect hyperplane separation happens at the point $p/n = 1/2$. Recall that Theorems \ref{Thm1}--\ref{Thm2} establish that the conventional GLM p-values can become invalid when $p \sim n^{2/3}$. We set $p = [n^{\alpha_0}]$ with $\alpha_0$ in the grid $\{2/3 - 4\delta, \cdots , 2/3 - \delta,
2/3, 2/3 + \delta, \cdots , 2/3 + 4\delta, (\log(n) - \log(2)) / \log(n)\}$  for $\delta = 0.05$. For example 3, we pick $s$ signals uniformly at random among all but the first components, where a random half of them are chosen as  $3$ and the other half are set as $-3$.


The goal of the simulation examples is to investigate empirically when the conventional GLM p-values could break down in diverging dimensions. When the asymptotic theory for the MLE in (\ref{011}) holds, the conventional p-values would be valid and distributed uniformly on the interval $[0, 1]$ under the null hypothesis. Note that the first covariate $\bx_1$ is a null variable in each simulation example. Thus in each replication, we calculate the conventional $p$-value for testing the null hypothesis $H_0: \beta_{0,1} = 0$. To check the validity of these p-values, we further test their uniformity.


For each simulation example, we first calculate the p-values for a total of  $1,000$ replications as described above and then test the uniformity of these $1,000$ p-values using, for example, the Kolmogorov--Smirnov (KS) test \cite{Kolmogorov1933, Smirnov1948} and the Anderson--Darling (AD) test \cite{AndersonDarling1952, AndersonDarling1954}. We repeat this procedure $1,000$ times to obtain a final set of $1,000$ new p-values from each of these two uniformity tests. 
Specifically, the KS and AD test statistics for testing the uniformity on $[0, 1]$ are defined as
\[
\text{KS}  =  \sup _{x \in [0, 1]} | F_{m} (x) - x | \ \text{ and } \
\text{AD}  =  m \int _{0} ^{1} \frac{[F_{m}(x) - x]^{2}}{x(1-x)} \, dx,
\]
respectively, where $F_{m} (x) = m^{-1} \sum _{i = 1} ^{m} I_{(-\infty, x]} (x_{i})$ is the empirical distribution function for a given sample $ \{ x_{i} \} _{i=1} ^{m}$.


\begin{figure}
\centering
\begin{subfigure}{.5\textwidth}
  \centering
  \includegraphics[width=0.99\linewidth]{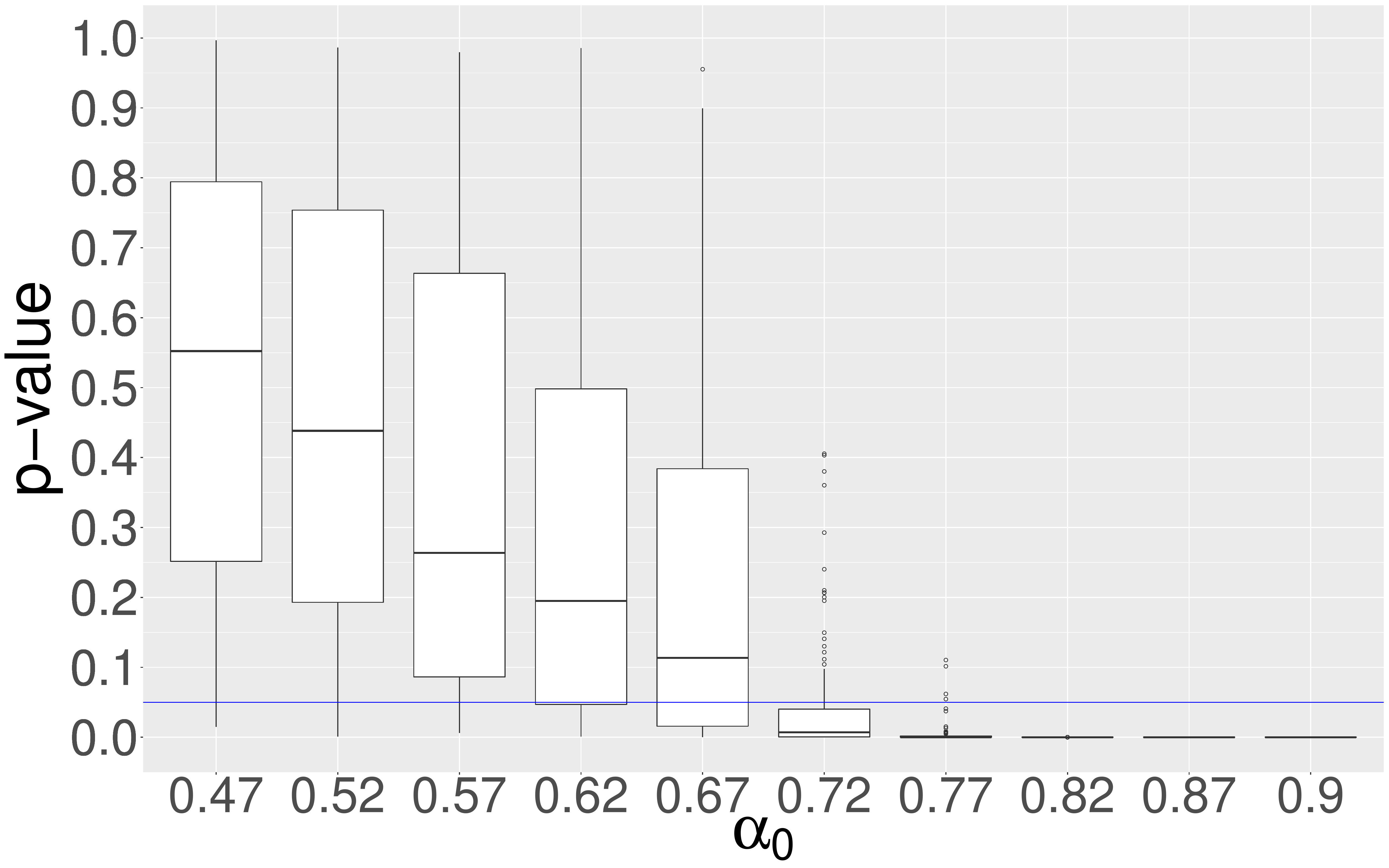}
  \caption{KS test}
\end{subfigure}%
\begin{subfigure}{.5\textwidth}
  \centering
  \includegraphics[width=0.99\linewidth]{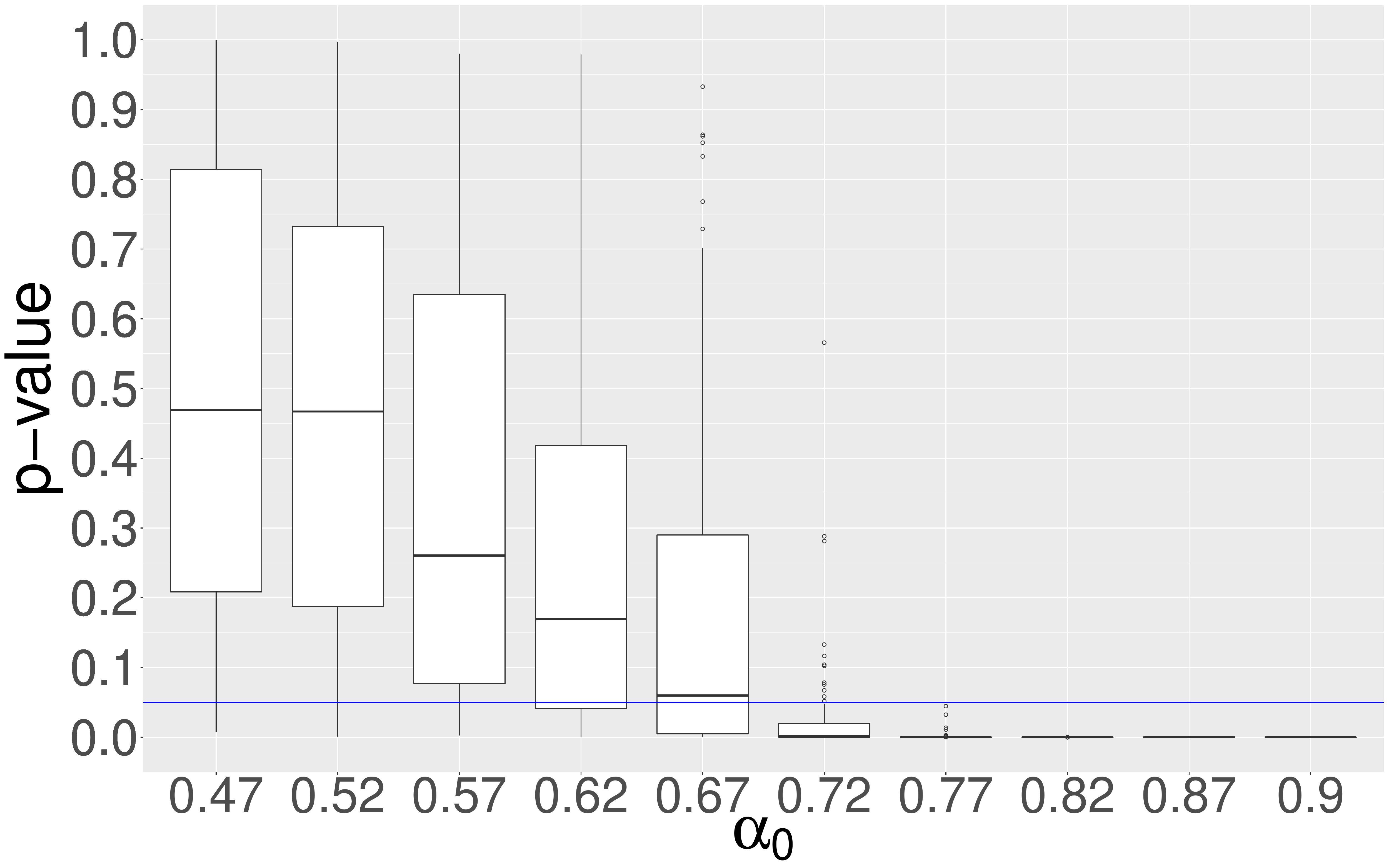}
  \caption{AD test}
\end{subfigure}
\caption{Results of KS and AD tests for testing the uniformity of GLM p-values in simulation example 1 for diverging-dimensional logistic regression model with uniform orthonormal design under global null. The vertical axis represents the p-value from the KS and AD tests, and the horizontal axis stands for the growth rate $\alpha_0$ of dimensionality $p = [n^{\alpha_0}]$.}
\label{fig1}
\end{figure}

\begin{figure}
\centering
\begin{subfigure}{.5\textwidth}
  \centering
  \includegraphics[width=0.99\linewidth]{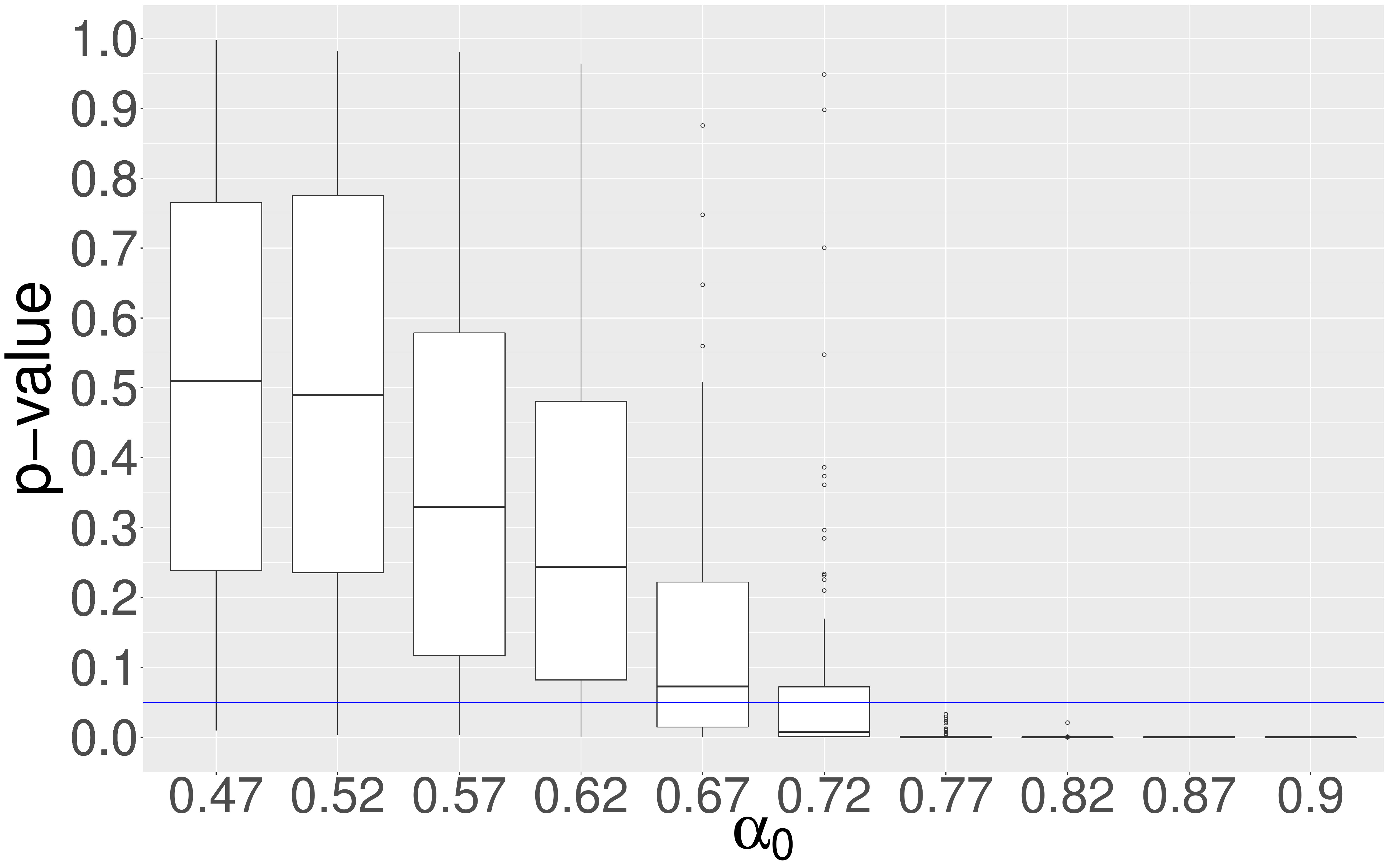}
  \caption{KS test for $\rho = 0.5$}
\end{subfigure}%
\begin{subfigure}{.5\textwidth}
  \centering
  \includegraphics[width=0.99\linewidth]{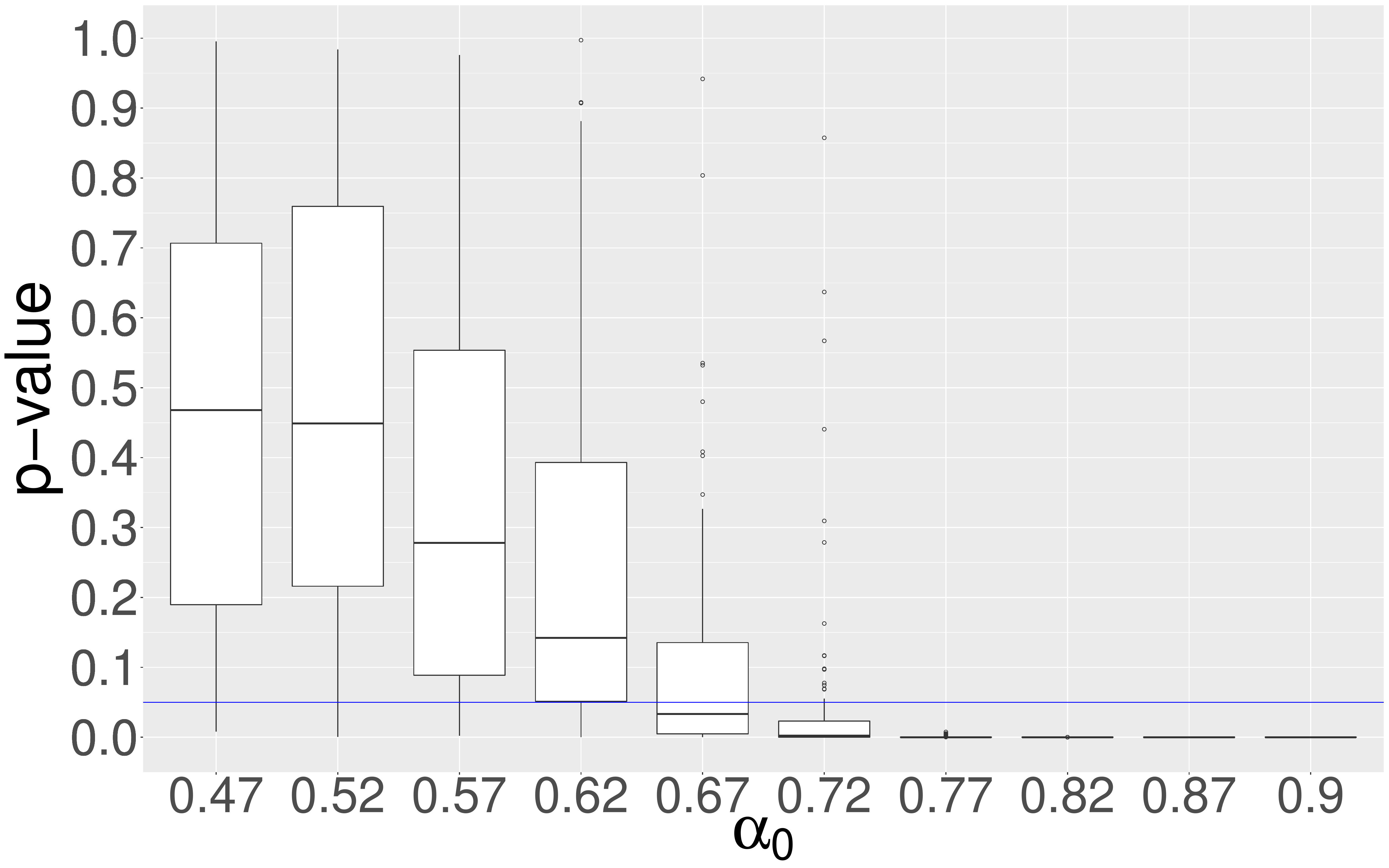}
  \caption{AD test for $\rho = 0.5$}
\end{subfigure}
\\
\begin{subfigure}{.5\textwidth}
  \centering
  \includegraphics[width=0.99\linewidth]{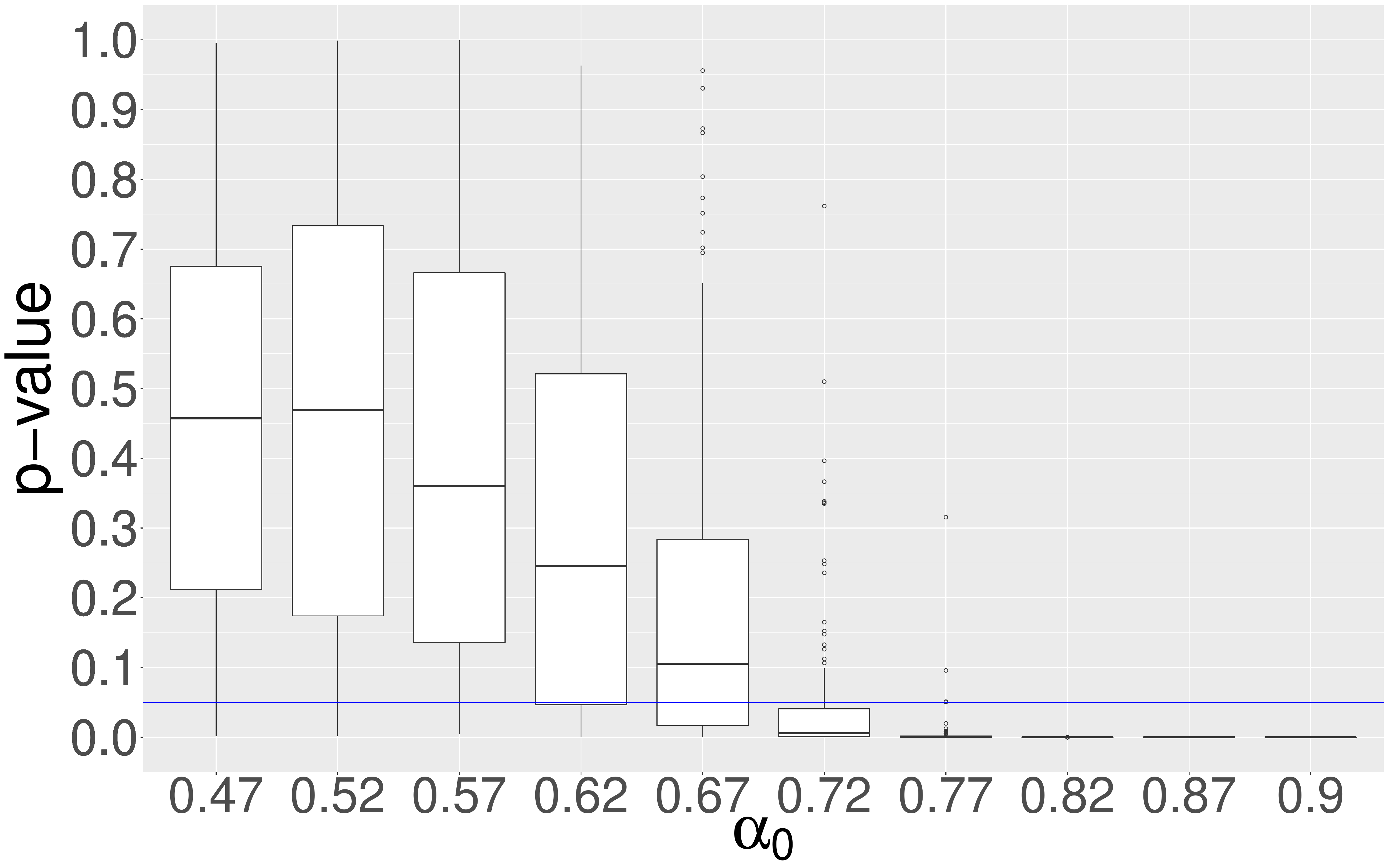}
  \caption{KS test for $\rho = 0.8$}
\end{subfigure}%
\begin{subfigure}{.5\textwidth}
  \centering
  \includegraphics[width=0.99\linewidth]{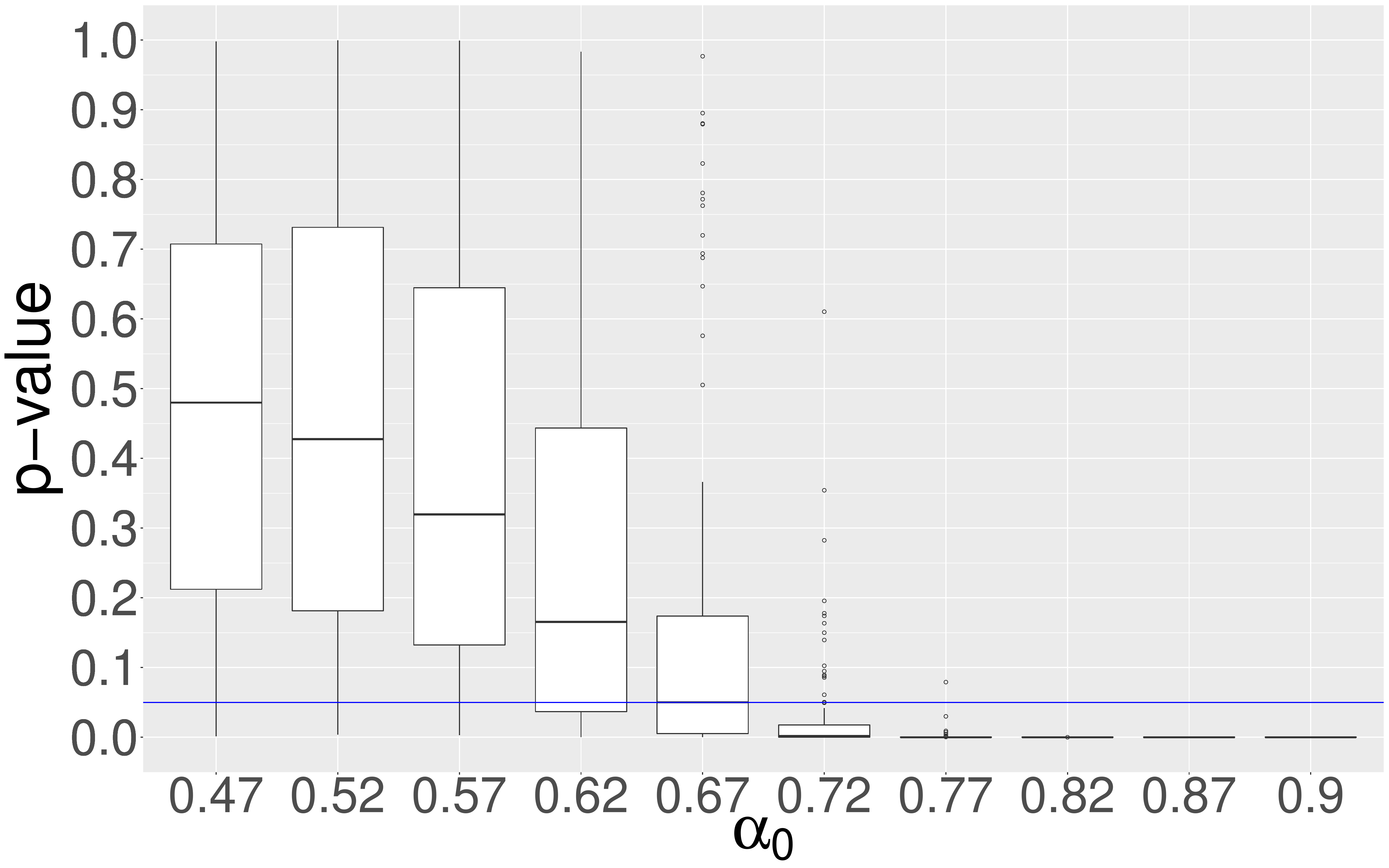}
  \caption{AD test for $\rho = 0.8$}
\end{subfigure}
\caption{Results of KS and AD tests for testing the uniformity of GLM p-values in simulation example 2 for diverging-dimensional logistic regression model with correlated Gaussian design under global null for varying correlation level $\rho$. The vertical axis represents the p-value from the KS and AD tests, and the horizontal axis stands for the growth rate $\alpha_0$ of dimensionality $p = [n^{\alpha_0}]$.}
\label{fig2}
\end{figure}

\begin{figure}
\centering
\begin{subfigure}{.5\textwidth}
  \centering
  \includegraphics[width=0.99\linewidth]{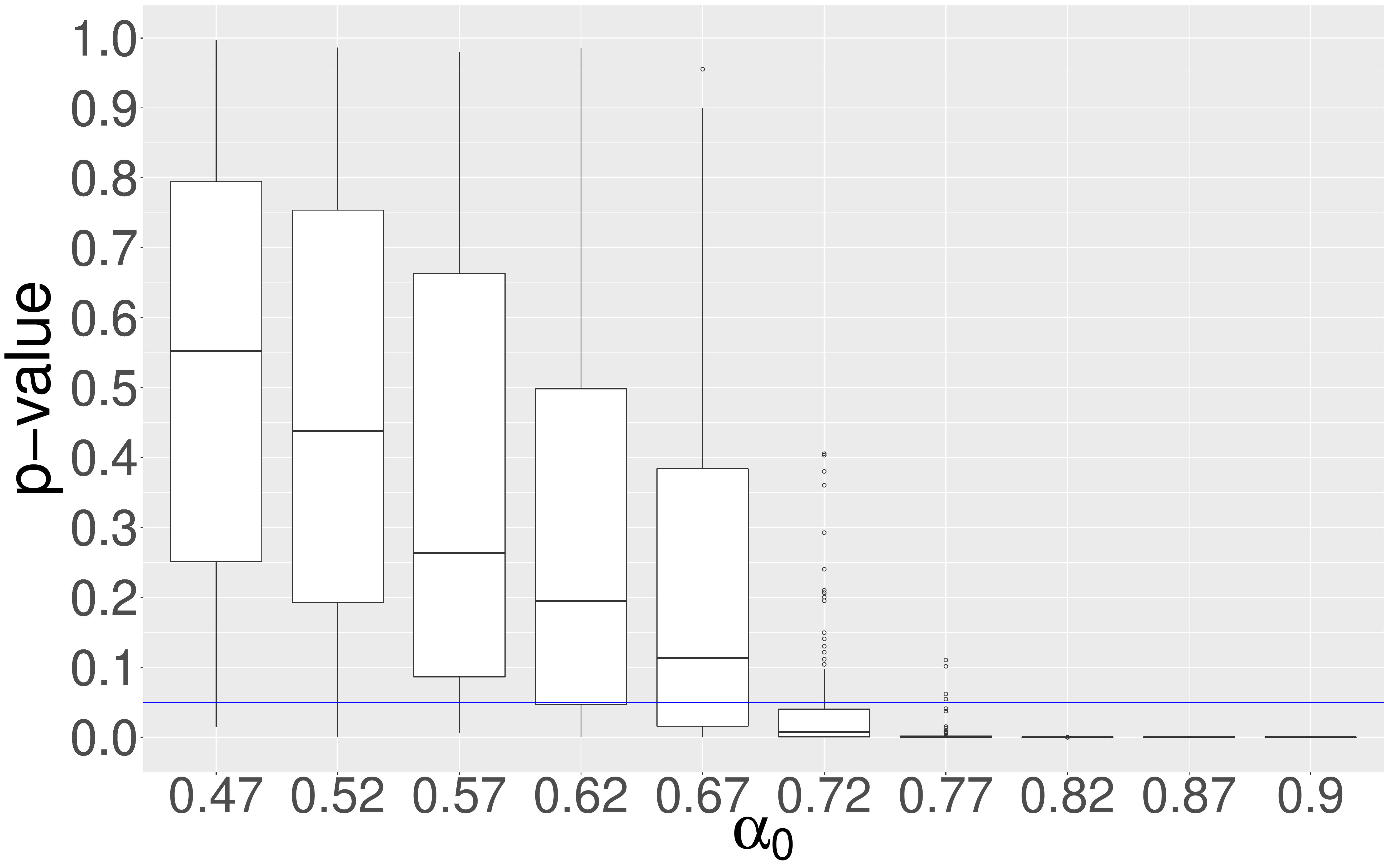}
  \caption{KS test for $s = 0$}
\end{subfigure}%
\begin{subfigure}{.5\textwidth}
  \centering
  \includegraphics[width=0.99\linewidth]{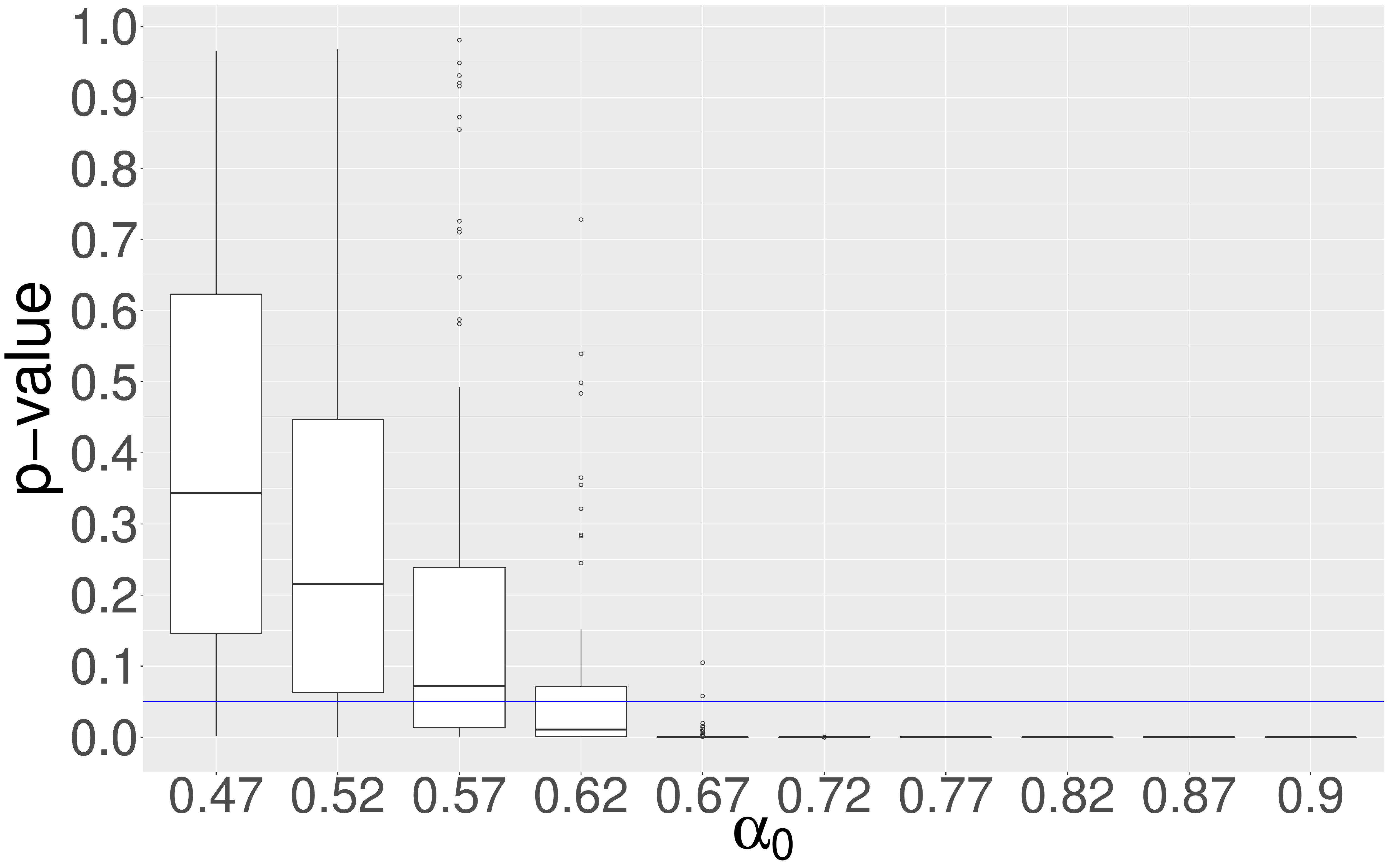}
  \caption{KS test for $s = 2$}
\end{subfigure}
\caption{Results of KS test for testing the uniformity of GLM p-values in simulation example 3 for diverging-dimensional logistic regression model with uncorrelated Gaussian design under global null for varying sparsity $s$. The vertical axis represents the p-value from the KS test, and the horizontal axis stands for the growth rate $\alpha_0$ of dimensionality $p = [n^{\alpha_0}]$.}
\label{fig3}
\end{figure}

\subsection{Testing results} \label{Sec4.2}

For each simulation example, we apply both KS and AD tests to verify the asymptotic theory for the MLE in (\ref{011}) by testing the uniformity of conventional p-values at significance level $0.05$. As mentioned in Section \ref{Sec4.1}, we end up with two sets of $1000$ new p-values from the KS and AD tests. Figures \ref{fig1}--\ref{fig3} depict the boxplots of the p-values obtained from both KS and AD tests for simulation examples 1--3, respectively. In particular, we observe that the numerical results shown in Figures \ref{fig1}--\ref{fig2} for examples 1--2 are in line with our theoretical results established in Theorems \ref{Thm1}--\ref{Thm2}, respectively, for diverging-dimensional logistic regression model under global null that the conventional p-values break down  when $p \sim n^{\alpha_0}$ with $\alpha_0 = 2/3$. Figure \ref{fig3} for example 3 examines the breakdown point of p-values with varying sparsity $s$. It is interesting to see that the breakdown point shifts even earlier when $s$ increases as suggested in the discussions in Section \ref{Sec3.2}. The results from the AD test are similar so we present only the results from the KS test for simplicity.

\section{Discussions} \label{Sec5}
In this paper we have provided characterizations of p-values in nonlinear GLMs with diverging dimensionality. The major findings are that the conventional p-values can remain valid when $p=o(n^{1/2})$, but can become invalid much earlier in nonlinear models of GLMs than in linear models, where the latter case can allow for $p = o(n)$. In particular, our theoretical results pinpoint the 
breakdown point of $p\sim n^{2/3}$ for p-values in diverging-dimensional logistic regression model under global null with uniform orthonormal design and correlated Gaussian design, as evidenced in the numerical results. It would be interesting to investigate such a phenomenon for more general class of random design matrices.


The problem of identifying the breakdown point of p-values becomes even more complicated and challenging when we move away from the setting of global null. Our technical analysis suggests that the breakdown point $p \sim n^{\alpha_0}$ can shift even earlier with $\alpha_0$ ranging between $1/2$ and $2/3$. But the exact breakdown point can depend upon the number of signals $s$, the signal magnitude, and the correlation structure among the covariates in a rather complicated fashion. Thus more delicate mathematical analysis is needed to obtain the exact relationship. We leave such a problem for future investigation. Moving beyond the GLM setting will further complicate the theoretical analysis.

As we routinely produce p-values using algorithms, the phenomenon of nonuniformity of p-values occurring early in diverging dimensions unveiled in the paper poses useful cautions to researchers and practitioners when making decisions in real applications using results from p-value based methods. For instance, when testing the joint significance of covariates in diverging-dimensional nonlinear models, the effective sample size requirement should be checked before interpreting the testing results. Indeed, statistical inference in general high-dimensional nonlinear models is particularly challenging since obtaining accurate p-values is generally uneasy. One possible route is to bypass the use of p-values in certain tasks including the false discovery rate (FDR) control; see, for example, \cite{BarberCandes2015, CandesFanJansonLv2016} for some initial efforts made along this line.

%
%

\appendix

\section{Proofs of main results} \label{SecA}
We provide the detailed proofs of Theorems \ref{Thm3}--\ref{Thm2} in this Appendix.

\subsection{Proof of Theorem \ref{Thm3}} \label{SecA.1}
To ease the presentation, we split the proof into two parts, where the first part locates the MLE $\hbbeta$ in an asymptotically shrinking neighborhood $\mathcal{N}_0$ of the true regression coefficient vector $\bbeta_0$ with significant probability and the second part further establishes its asymptotic normality.

\smallskip

\textbf{Part 1:} Existence of a solution to score equation (\ref{021}) in $\mathcal{N}_0$ under Condition \ref{con1} and probability bound (\ref{137}). For simplicity, assume that the design matrix $\bX$ is rescaled columnwise such that $\|\bx_j\|_2 = \sqrt{n}$ for each $1 \leq j \leq p$.
Consider an event
\begin{equation} \label{003}
\mathcal{E} = \left\{\left\|\bxi\right\|_\infty \leq c_1^{-1/2} \sqrt{n \log n}\right\},
\end{equation}
where $\bxi = (\xi_1, \cdots, \xi_p)\t = \bX\t [\by - \bmu(\btheta_0)]$. Note that for unbounded responses, the assumption of $\max_{j = 1}^p \|\bx_j\|_\infty < c_1^{1/2} \{n/(\log n)\}^{1/2}$ in Condition \ref{con1} entails that $c_1^{-1/2}
\sqrt{\log n} < \min_{j = 1}^p \{\|\bx_j\|_2/\|\bx_j\|_\infty\}$. Thus by $\|\bx_j\|_2 = \sqrt{n}$, probability bound (\ref{137}), and Bonferroni's inequality, we deduce
\begin{align} \label{106}
P\left(\mathcal{E}\right) & \geq 1 - \sum_{j=1}^p P\left(|\xi_j| > c_1^{-1/2}
\sqrt{n \log n}\right) \\
& \geq 1 - 2 p n^{-1} = 1 - O\{n^{-(1 - \alpha_0)}\}, \nonumber
\end{align}
since $p = O(n^{\alpha_0})$ for some $\alpha_0 \in (0, \gamma)$ with $\gamma \in (0,1/2]$ by assumption. Hereafter we condition on the event $\mathcal{E}$ defined in (\ref{003}) which holds with significant probability.

We will show that for sufficiently large $n$, the score equation (\ref{021}) has a solution in the neighborhood $\mathcal{N}_0$ which is a hypercube. Define two vector-valued functions
\[ \bgamma(\bbeta) = (\gamma_1(\bbeta), \cdots, \gamma_p(\bbeta))\t = \bX\t \bmu(\bX \bbeta) \]
and
\[ \bPsi(\bbeta) = \bgamma(\bbeta) - \bgamma(\bbeta_0) - \bxi, \quad \bbeta \in \mathbb{R}^p. \]
Then equation (\ref{021}) is equivalent to $\bPsi(\bbeta) = \bzero$.  We need to show that the latter has a solution inside the hypercube $\mathcal{N}_0$. To this end, applying a second order Taylor expansion of $\bgamma(\bbeta)$ around $\bbeta_0$ with the Lagrange remainder term componentwise leads to
\begin{equation} \label{110}
\bgamma(\bbeta) =
\bgamma(\bbeta_0) +  \bX\t
\Sig\left(\btheta_0\right) \bX (\bbeta - \bbeta_0) + \br,
\end{equation}
where $\br = (r_1, \cdots, r_p)\t$ and for each $1 \leq j \leq p$,
\[ r_j = \frac{1}{2} \left(\bbeta - \bbeta_0\right)\t \nabla^2 \gamma_{j}(\bbeta_j) \left(\bbeta - \bbeta_0\right) \]
with $\bbeta_j$ some $p$-dimensional vector lying on the line segment joining $\bbeta$ and $\bbeta_0$. It follows from (\ref{009}) in Condition \ref{con1} that
\begin{align} \label{111}
\left\|\br\right\|_\infty & \leq \max_{\bdelta \in \mathcal{N}_0} \max_{j = 1}^p \frac{1}{2} \lambda_{\max}\left[\bX\t \diag\left\{\left|\bx_j\right| \circ \left|\bmu''\left(\bX \bdelta\right)\right|\right\} \bX\right] \left\|\bbeta - \bbeta_0\right\|_2^2 \\
\nonumber
&  = O\left\{p n^{1 - 2 \gamma} (\log n)^2\right\}.
\end{align}

Let us define another vector-valued function
\begin{equation} \label{112}
\overline{\bPsi}(\bbeta) \equiv  \left[\bX\t
\Sig\left(\btheta_0\right) \bX\right]^{-1} \bPsi(\bbeta) =
\bbeta - \bbeta_0 + \bu,
\end{equation}
where  $\bu = -[\bX\t \Sig(\btheta_0) \bX]^{-1}
(\bxi - \br)$. It follows from (\ref{003}),
(\ref{111}), and (\ref{007}) in Condition \ref{con1} that for any
$\bbeta \in \mathcal{N}_0$,
\begin{align} \label{140}
\left\|\bu\right\|_\infty
& \leq  \left\|\left[\bX\t \Sig\left(\btheta_0\right) \bX\right]^{-1}\right\|_\infty \left(\|\bxi\|_\infty + \|\br\|_\infty\right) \\
& = O\left[b_n n^{-1/2} \sqrt{\log n} + b_n p n^{- 2 \gamma} (\log n)^2\right]. \nonumber
\end{align}
By the assumptions of $p = O(n^{\alpha_0})$ with constant $\alpha_0 \in (0, \gamma)$ and $b_n = o\{\min(n^{1/2 -\gamma} \sqrt{\log n}, n^{2\gamma - \alpha_0 - 1/2}/ (\log n)^2\}$, we have
\[
   \left\|\bu\right\|_\infty  = o(n^{-\gamma} \log n). \]
Thus in light of (\ref{112}), it holds for large enough $n$ that when $(\bbeta - \bbeta_0)_j = n^{-\gamma} \sqrt{\log n}$,
\begin{equation} \label{114}
\overline{\Psi}_j(\bbeta) \geq n^{-\gamma} \sqrt{\log n} - \left\|\bu\right\|_\infty  \geq 0,
\end{equation}
and when $(\bbeta - \bbeta_0)_j = -n^{-\gamma} \sqrt{\log n}$,
\begin{equation} \label{115}
\overline{\Psi}_j(\bbeta) \leq -n^{-\gamma} \sqrt{\log n} + \left\|\bu\right\|_\infty \leq 0,
\end{equation}
where $\overline{\bPsi}(\bbeta) = (\overline{\Psi}_1(\bbeta), \cdots, \overline{\Psi}_p(\bbeta))\t$.

By the continuity of the vector-valued function $\overline{\bPsi}(\bbeta)$, (\ref{114}), and (\ref{115}), Miranda's existence theorem \cite{Vrahatis89} ensures that equation $\overline{\bPsi}(\bbeta) = \bzero$ has a solution $\hbbeta$ in $\mathcal{N}_0$. Clearly, $\hbbeta$ also solves equation $\bPsi(\bbeta) = \bzero$ in view of (\ref{112}). Therefore, we have shown that score equation (\ref{021}) indeed has a solution $\hbbeta$ in $\mathcal{N}_0$. The strict concavity of the log-likelihood function (\ref{002}) by assumptions for model (\ref{001}) entails that $\hbbeta$ is the MLE.

\smallskip

\textbf{Part 2:} Conventional asymptotic normality of the MLE $\hbbeta$.
Fix any $1 \leq j \leq p$. In light of (\ref{112}), we have $\hbbeta - \bbeta_0 = \bA_n^{-1}(\bxi - \br)$, which results in
\begin{equation} \label{006}
(\bA_n^{-1})_{jj}^{-1/2} (\hbeta_j - \beta_{0,j}) = (\bA_n^{-1})_{jj}^{-1/2} \be_j\t \bA_n^{-1} \bxi - (\bA_n^{-1})_{jj}^{-1/2} \be_j\t \bA_n^{-1} \br
\end{equation}
with $\be_j \in \mathbb{R}^p$ having one for the $j$th component and zero otherwise. Note that since the smallest and largest eigenvalues of $n^{-1} \bA_n$ are bounded away from $0$ and $\infty$ by Condition \ref{con2}, it is easy to show that  $(\bA_n^{-1})_{jj}^{-1/2}$ is of exact order $n^{1/2}$. In view of (\ref{140}), it holds on the event $\mathcal{E}$ defined in (\ref{003}) that
\begin{eqnarray*}
\left\|\bA_n^{-1} \br\right\|_\infty
& \leq  & \left\|\left[\bX\t \Sig\left(\btheta_0\right) \bX\right]^{-1}\right\|_\infty \|\br\|_\infty \\
& = & O\left[b_n p n^{- 2 \gamma} (\log n)^2\right] = o(n^{-1/2}),
\end{eqnarray*}
since $b_n = o\{n^{2\gamma - \alpha_0 - 1/2}/ (\log n)^2\}$ by assumption. This leads to
\begin{equation} \label{005}
(\bA_n^{-1})_{jj}^{-1/2} \be_j\t \bA_n^{-1} \br = O(n^{1/2}) \cdot o_P(n^{-1/2}) = o_P(1).
\end{equation}

It remains to consider the term $(\bA_n^{-1})_{jj}^{-1/2} \be_j\t \bA_n^{-1} \bxi = \sum_{i = 1}^n \eta_i$, where $\eta_i = (\bA_n^{-1})_{jj}^{-1/2} \be_j\t \bA_n^{-1} \bz_i [y_i - b'(\theta_{0,
i})]$. Clearly, the $n$ random variables $\eta_i$'s are
independent with mean 0 and
\[
\sum_{i = 1}^n \var \eta_i = (\bA_n^{-1})_{jj}^{-1} \be_j\t \bA_n^{-1} (\aphi \bA_n) \bA_n^{-1} \be_j = \aphi.
\]
It follows from Condition \ref{con2} and the Cauchy--Schwarz inequality that
\begin{align*}
\sum_{i = 1}^n E \left|\eta_i\right|^3 & =  \sum_{i = 1}^n \left|(\bA_n^{-1})_{jj}^{-1/2} \be_j\t \bA_n^{-1} \bz_i\right|^3 E \left|y_i - b'\left(\theta_{0, i}\right)\right|^3 \\
& = O(1) \sum_{i = 1}^n \left|(\bA_n^{-1})_{jj}^{-1/2} \be_j\t \bA_n^{-1} \bz_i\right|^3 \\
& \leq O(1) \sum_{i = 1}^n \left\|(\bA_n^{-1})_{jj}^{-1/2} \be_j\t \bA_n^{-1/2}\right\|_2^3 \left\|\bA_n^{-1/2} \bz_i\right\|_2^3 \\
& = O(1) \sum_{i = 1}^n \left(\bz_i\t \bA_n^{-1} \bz_i\right)^{3/2} = o(1).
\end{align*}
Thus an application of Lyapunov's theorem yields
\begin{equation} \label{008}
(\bA_n^{-1})_{jj}^{-1/2} \be_j\t \bA_n^{-1} \bxi
= \sum_{i = 1} ^n \eta _i \toD N(0, \aphi).
\end{equation}
By Slutsky's lemma, we see from (\ref{006})--(\ref{008}) that
\[
(\bA_n^{-1})_{jj}^{-1/2} (\hbeta_j - \beta_{0,j}) \toD N(0, \aphi),
\]
showing the asymptotic normality of each component $\hbeta_j$ of the MLE $\hbbeta$. This completes the proof of Theorem \ref{Thm3}.

\subsection{Proof of Theorem \ref{Thm1}} \label{SecA.2}
To prove the conclusion in Theorem \ref{Thm1}, we use the proof by contradiction. Let us make an assumption (A) that the asymptotic normality (\ref{011}) in Theorem \ref{Thm3} which has been proved to hold when $p = o(n^{1/2})$ continues to hold when $p \sim n^{\alpha_0}$ for  some constant $1/2 < \alpha_0 \leq 1$, where $\sim$ stands for asymptotic order. As shown in Section \ref{Sec3.1}, in the case of logistic regression under global null (that is, $\bbeta_0 = \bzero$) with deterministic rescaled orthonormal design matrix $\bX$ (in the sense of $n^{-1} \bX\t \bX = I_p$) the limiting distribution in (\ref{011}) by assumption (A) becomes
\begin{equation} \label{012}
2^{-1} n^{1/2} \hbeta_j \toD N(0, 1),
\end{equation}
where $\hbbeta = (\hbeta_1, \cdots, \hbeta_p)\t$ is the MLE.

Let us now assume that the rescaled random design matrix $n^{-1/2} \bX$ is uniformly distributed on the Stiefel manifold $V_p(\mathbb{R}^n)$ which can be thought of as the space of all $n \times p$ orthonormal matrices. Then it follows from (\ref{012}) that
\begin{equation} \label{013}
2^{-1} n^{1/2} \hbeta_j \toD N(0, 1) \text{ \ conditional on $\bX$}.
\end{equation}
Based on the limiting distribution in (\ref{013}), we can make two observations. First, it holds that
\begin{equation} \label{014}
2^{-1} n^{1/2} \hbeta_j \toD N(0, 1)
\end{equation}
unconditional on the design matrix $\bX$. Second, $\hbeta_j$ is asymptotically independent of the design matrix $\bX$, and so is the MLE $\hbbeta$.

Since the distribution of $n^{-1/2} \bX$ is assumed to be the Haar measure on the Stiefel manifold $V_p(\mathbb{R}^n)$, we have
\begin{equation} \label{015}
n^{-1/2} \bX \bQ \eqd n^{-1/2} \bX,
\end{equation}
where $\bQ$ is any fixed $p \times p$ orthogonal matrix and $\eqd$ stands for equal in distribution. Recall that the MLE $\hbbeta$ solves the score equation (\ref{021}), which is in turn equivalent to equation
\begin{equation} \label{016}
\bQ\t \bX\t [\by - \bmu(\bX \bbeta)] = \bzero
\end{equation}
since $\bQ$ is orthogonal. We now use the fact that the model is under global null which entails that the response vector $\by$ is independent of the design matrix $\bX$. Combining this fact with (\ref{015})--(\ref{016}) yields
\begin{equation} \label{017}
\bQ\t \hbbeta \eqd \hbbeta
\end{equation}
by noting that $\bX \bbeta = (\bX \bQ) (\bQ\t \bbeta)$. Since the distributional identity (\ref{017}) holds for any fixed $p \times p$ orthogonal matrix $\bQ$, we conclude that the MLE $\hbbeta$ has a spherical distribution on $\mathbb{R}^p$. It is a well-known fact that all the marginal characteristic functions of a spherical distribution have the same generator. Such a fact along with (\ref{014}) entails that
\begin{equation} \label{018}
2^{-1} n^{1/2} \hbbeta \text{ is asymptotically close to } N(\bzero, I_p).
\end{equation}

To simplify the exposition, let us now make the asymptotic limit exact and assume that
\begin{equation} \label{019}
\hbbeta \sim N(\bzero, 4n^{-1} I_p) \text{ and is independent of } \bX.
\end{equation}
The remaining analysis focuses on the score equation (\ref{021}) which is solved exactly by the MLE $\hbbeta$, that is,
\begin{equation} \label{020}
\bX\t [\by - \bmu(\bX \hbbeta)] = \bzero,
\end{equation}
which leads to
\begin{equation} \label{022}
\bxi \equiv n^{-1/2} \bX\t [\by - \bmu(\bzero)] = n^{-1/2} \bX\t [\bmu(\bX \hbbeta) - \bmu(\bzero)] \equiv \bet.
\end{equation}
Let us first consider the random variable $\bxi$ defined in (\ref{022}). Note that $2[\by - \bmu(\bzero)]$ has independent and identically distributed (i.i.d.) components each taking value $1$ or $-1$ with equal probability $1/2$, and is independent of $\bX$. Thus since $n^{-1/2} \bX$ is uniformly distributed on the Stiefel manifold $V_p(\mathbb{R}^n)$, it is easy to see that
\begin{equation} \label{023}
\bxi = n^{-1/2} \bX\t [\by - \bmu(\bzero)] \eqd 2^{-1}n^{-1/2} \bX\t \bone,
\end{equation}
where $\bone \in \mathbb{R}^n$ is a vector with all components being one. Using similar arguments as before, we can show that $\bxi$ has a spherical distribution on $\mathbb{R}^p$. Thus the joint distribution of $\bxi$ is determined completely by the marginal distribution of $\bxi$. For each $1 \leq j \leq p$, denote by $\xi_j$ the $j$th component of $\bxi = 2^{-1}n^{-1/2} \bX\t \bone$ using the distributional representation in (\ref{023}). Let $\bX = (\bx_1, \cdots, \bx_p)$ with each $\bx_j \in \mathbb{R}^n$. Then we have
\begin{equation} \label{024}
\xi_j = 2^{-1}n^{-1/2} \bx_j\t \bone \eqd 2^{-1}(n^{1/2}/\|\widetilde{\bx}_j\|_2) n^{-1/2} \widetilde{\bx}_j\t \bone,
\end{equation}
where $\widetilde{\bx}_j \sim N(\bzero, 4^{-1}I_n)$. It follows from (\ref{024}) and the concentration phenomenon of Gaussian measures that each $\xi_j$ is asymptotically close to $N(0, 4^{-1})$ and thus consequently $\bxi$ is asymptotically close to $N(\bzero, 4^{-1} I_p)$. \textit{A key fact (i) for the finite-sample distribution of $\bxi$ is that the standard deviation of each component $\xi_j$ converges to $1/2$ at rate $O_P(n^{-1/2})$ that does not depend upon the dimensionality $p$ at all.}

We now turn our attention to the second term $\bet$ defined in (\ref{022}). In view of (\ref{019}) and the fact that $n^{-1/2} \bX$ is uniformly distributed on the Stiefel manifold $V_p(\mathbb{R}^n)$, we can show that with significant probability,
\begin{equation} \label{025}
\|\bX \hbbeta\|_\infty \leq o(1)
\end{equation}
for $p \sim n^{\alpha_0}$ with $\alpha_0 < 1$. The uniform bound in (\ref{025}) enables us to apply the mean value theorem for the vector-valued function $\bet$ around $\bbeta_0 = \bzero$, which results in
\begin{align} \label{026}
\bet & = n^{-1/2} \bX\t [\bmu(\bX \hbbeta) - \bmu(\bzero)] = 4^{-1}n^{-1/2} \bX\t \bX \hbbeta + \br \\
& = 4^{-1}n^{1/2} \hbbeta + \br \nonumber
\end{align}
since $n^{-1/2} \bX$ is assumed to be orthonormal, where
\begin{equation} \label{028}
\br = n^{-1/2} \bX\t \left\{\int_0^1\left[\Sig(t \bX \hbbeta) - 4^{-1}I_n\right] dt\right\} \bX \hbbeta.
\end{equation}
Here, the remainder term $\br = (r_1, \cdots, r_p)\t \in \mathbb{R}^p$ is stochastic and each component $r_j$ is generally of order $O_P\{p^{1/2}n^{-1/2}\}$ in light of (\ref{019}) when the true model may deviate from the global null case of $\bbeta_0 = \bzero$.

Since our focus in this theorem is the logistic regression model under the global null, we can in fact claim that each component $r_j$ is generally of order $O_P\{pn^{-1}\}$, which is a better rate of convergence than the one mentioned above thanks to the assumption of $\bbeta_0 = \bzero$. To prove this claim, note that the variance function $b''(\theta)$ is symmetric in $\theta \in \mathbb{R}$ and takes the maximum value $1/4$ at $\theta = 0$. Thus in view of (\ref{025}), we can show that with significant probability,
\begin{equation} \label{027}
4^{-1}I_n - \Sig(t \bX \hbbeta) \geq c \diag\{(t \bX \hbbeta) \circ (t \bX \hbbeta)\} = c t^2 \diag\{(\bX \hbbeta) \circ (\bX \hbbeta)\}
\end{equation}
for all $t \in [0, 1]$, where $c > 0$ is some constant and $\geq$ stands for the inequality for positive semidefinite matrices. Moreover, it follows from (\ref{019}) and the fact that $n^{-1/2} \bX$ is uniformly distributed on the Stiefel manifold $V_p(\mathbb{R}^n)$ that with significant probability, all the $n$ components of $\bX \hbbeta$ are concentrated in the order of $p^{1/2} n^{-1/2}$. This result along with (\ref{027}) and the fact that $n^{-1} \bX\t \bX = I_{p}$ 
entails that with significant probability,
\begin{align} \label{029}
n^{-1/2} & \bX\t \left\{\int_0^1\left[4^{-1}I_n - \Sig(t \bX \hbbeta)\right] dt\right\} \bX  \\
& \geq n^{-1/2} \bX\t \left\{\int_0^1 c_* t^2 p n^{-1} dt\right\} \bX \nonumber \\
& = 3^{-1} c_* p n^{-3/2} \bX\t \bX 
= 3^{-1} c_* p n^{-1/2} I_p, \nonumber
\end{align}
where $c_* > 0$ is some constant. Thus combining (\ref{028}), (\ref{029}), and (\ref{019}) proves the above claim.

We make two important observations about the remainder term $\br$ in (\ref{026}). First, $\br$ has a spherical distribution on $\mathbb{R}^p$. This is because by (\ref{026}) and (\ref{022}) it holds that
\[ \br = \bet - 4^{-1} n^{1/2} \hbbeta = \bxi - 4^{-1}n^{1/2} \hbbeta, \]
which has a spherical distribution on $\mathbb{R}^p$. Thus the joint distribution of $\br$ is determined completely by the marginal distribution of $\br$. Second, for the nonlinear setting of logistic regression model, the appearance of the remainder term $\br$ in (\ref{026}) is due solely to \textit{the nonlinearity of the mean function} $\bmu(\cdot)$, and we have shown that each component $r_j$ can indeed achieve the worst-case order $pn^{-1}$ in probability. For each $1 \leq j \leq p$, denote by $\eta_j$ the $j$th component of $\bet$. Then in view of (\ref{019}) and (\ref{026}), \textit{a key fact (ii) for the finite-sample distribution of $\bet$ is that the standard deviation of each component $\eta_j$ converges to $1/2$ at rate $O_P\{pn^{-1}\}$ that generally does depend upon the dimensionality $p$.} 

Finally, we are ready to compare the two random variables $\bxi$ and $\bet$ on the two sides of equation (\ref{022}). Since equation (\ref{022}) is a distributional identity in $\mathbb{R}^p$, naturally the square root of the sum of $\var{\xi_j}$'s and the square root of the sum of $\var{\eta_j}$'s are expected to converge to the common value $2^{-1} p^{1/2}$ at rates that are asymptotically negligible. However, the former has rate $p^{1/2} O_P(n^{-1/2}) = O_P\{p^{1/2} n^{-1/2}\}$, whereas the latter has rate $p^{1/2} O_P\{pn^{-1}\} = O_P\{p^{3/2} n^{-1}\}$. A key consequence is that when $p \sim n^{\alpha_0}$ for  some constant $2/3 \leq \alpha_0 < 1$, there is a profound difference between the two asymptotic rates in that the former rate is $O_P\{n^{-(1-\alpha_0)/2}\} = o_P(1)$, while the latter rate becomes $O_P\{n^{3\alpha_0/2 - 1}\}$ which is now asymptotically diverging or nonvanishing. Such an intrinsic asymptotic difference is, however, prohibited by the distributional identity (\ref{022}) in $\mathbb{R}^p$, which results in a contradiction. Therefore, we have now argued that assumption (A) we started with for $2/3 \leq \alpha_0 < 1$ must be false, that is, the asymptotic normality (\ref{011}) which has been proved to hold when $p = o(n^{1/2})$ generally would not continue to hold when $p \sim n^{\alpha_0}$ with constant $2/3 \leq \alpha_0 \leq 1$. In other words, we have proved the invalidity of the conventional GLM p-values in this regime of diverging dimensionality, which concludes the proof of Theorem \ref{Thm1}.

\subsection{Proof of Theorem \ref{Thm2}} \label{SecA.3}
By assumption, $\bX \sim N (\bzero, I_{n} \otimes \Sig )$ with covariance matrix $\Sig$ nonsingular. Let us first make a useful observation. For the general case of nonsingular covariance matrix $\Sig$, we can introduce a change of variable by letting $\widetilde{\bbeta} = \Sig ^{1/2} \bbeta $ and correspondingly $\widetilde{\bX} = \bX \Sig ^{-1/2}$. Clearly, $\widetilde{\bX} \sim N (\bzero, I_{n} \otimes I_p)$ and the MLE for the transformed parameter vector $\widetilde{\bbeta}$ is exactly $\Sig ^{1/2} \hbbeta $, where $\hbbeta$ denotes the MLE under the original design matrix $\bX$. Thus to show the breakdown point of the conventional asymptotic normality of the MLE, it suffices to focus on the specific case of $\bX \sim N (\bzero, I_{n} \otimes I_p)$.

%
%

Hereafter we assume that $\bX \sim N (\bzero, I_{n} \otimes I_p)$ with $p = o(n)$. The rest of the arguments are similar to those in the proof of Theorem \ref{Thm1} in Section \ref{SecA.2} except for some modifications needed for the case of Gaussian design. Specifically, for the case of logistic regression model under global null (that is, $\bbeta_0 = \bzero$), the limiting distribution in (\ref{011}) becomes
\begin{equation} \label{eqn:lim_assump}
2^{-1} n^{1/2} \hbeta_j \toD N(0, 1),
\end{equation}
since $n^{-1} \bX\t \bX \rightarrow I_{p} $ almost surely in spectrum and thus $4^{-1} n (\bA_n^{-1})_{jj} \rightarrow 1 $ in probability as $n \rightarrow \infty$. Here, we have used a claim that both the largest and smallest eigenvalues of $n^{-1} \bX\t \bX$ converge to $1$ almost surely as $n \rightarrow \infty$ for the case of $p = o(n)$, which can be shown by using the classical results from random matrix theory (RMT) \cite{Geman1980, Silverstein1985, Bai1999}.

Note that since $\bX \sim N ( 0, I_{n} \otimes I_{p} )$, it holds that
\begin{equation} \label{eqn:inv_distr}
n^{-1/2} \bX \bQ \eqd
 n^{-1/2} \bX,
\end{equation}
where $\bQ$ is any fixed $p \times p$ orthogonal matrix and $\eqd$ stands for equal in distribution. By $\bX \sim N ( 0, I_{n} \otimes I_{p} )$, it is also easy to see that
\begin{equation} \label{eqn:xi}
\bxi = n^{-1/2} \bX\t [\by - \bmu(\bzero)] \eqd 2^{-1}n^{-1/2} \bX\t \bone,
\end{equation}
where $\bone \in \mathbb{R}^n$ is a vector with all components being one. In view of (\ref{019}) and the assumption of $\bX \sim N ( 0, I_{n} \otimes I_{p} )$, we can show that with significant probability,
\begin{equation} \label{eqn:xbeta_order}
\|\bX \hbbeta\|_\infty \leq o(1)
\end{equation}
for $p \sim n^{\alpha_0}$ with constant $\alpha_0 < 1$. It holds further that with significant probability, all the $n$ components of $\bX \hbbeta$ are concentrated in the order of $p^{1/2} n^{-1/2}$. This result along with (\ref{027}) and the fact that $n^{-1} \bX\t \bX \rightarrow I_{p} $ almost surely in spectrum entails that with asymptotic probability one, 
\begin{align} \label{eqn:variance2}
n^{-1/2} & \bX\t \left\{\int_0^1\left[4^{-1}I_n - \Sig(t \bX \hbbeta)\right] dt\right\} \bX \\
& \geq n^{-1/2} \bX\t \left\{\int_0^1 c_* t^2 p n^{-1} dt\right\} \bX \nonumber \\
& = 3^{-1} c_* p n^{-3/2} \bX\t \bX \rightarrow 3^{-1} c_* p n^{-1/2} I_p, \nonumber
\end{align}
where $c_* > 0$ is some constant. This completes the proof of Theorem \ref{Thm2}.

\bibliographystyle{imsart-nameyear}
\bibliography{references}

\end{document}